# Broadband Polarization-Independent Achromatic Metalenses with Unintuitively-Designed Random-Shaped Meta-Atoms

Xiaojie Zhang, Haiyang Huang, Xuexue Guo, Xingwang Zhang, Yao Duan, Xi Chen, Shengyuan Chang, Yimin Ding, Xingjie Ni[*]

Department of Electrical Engineering, Pennsylvania State University, University Park, PA 16802, USA.

* Correspondence to: xingjie@psu.edu

# Abstract

Metasurface lenses, namely metalenses, are ultrathin planar nanostructures that are capable of manipulating the properties of incoming light and imparting lens-like wavefront to the output. Although they have shown promising potentials for the future miniaturization of optics, the chromatic aberration inherited from their diffractive nature plagues them towards many practical applications. Current solutions for creating achromatic metalenses usually require searching through a large number of meta-atoms to find designs that fulfill not only phase but phase dispersion requirements, which leads to intensive design efforts. Besides, most designs are based on regular-shaped antennas driven by the designers' intuition and experience, hence only cover a limited design space. Here, we present an inverse design approach that efficiently produces meta-atoms with unintuitive geometries required for broadband achromatic metalenses. We restricted the generated shapes to hold four-fold reflectional symmetry so that the resulting metalenses are polarization insensitive. In addition, meta-atoms generated by our method inheritably have round edges and corners, which make them nanofabrication-friendly. Our experimental characterization shows that our metalenses exhibit superior performance over a broad bandwidth of 465 nm in the near-infrared regime. Our method offers a fast and efficient way of designing high-performance achromatic metalenses and sheds new insights for unintuitive design of other metaphotonic devices.

# Introduction

Metalens is a flat optical component, accomplished with an artificial subwavelength nanostructured interface that manipulates light by spatially variant nanoantennas (i.e. meta-atoms) [1–4]. Recently, the metalenses have attracted considerable interests due to their great potential for future high-performance portable or wearable optical devices and systems with small footprints and light weights [5–7]. In particular, achromatic metalenses that compensates chromatic aberrations within a wavelength range of interest are highly desirable in many practical applications such as microscopes, digital cameras, and 3D displays [8–13]. Assorted methods have been investigated to compensate the chromatic aberration. For example, multilayer metasurfaces can realize multiwavelength achromatic metalens where independent control over phase at different wavelengths is achieved at different layers [14]. Multiwavelength achromatic metalenses can be also realized using interleaved metasurfaces that are created by spatially distributing meta-atoms from several chromatic metalenses working at different wavelengths, respectively [15]. exploiting geometric phase combining with dispersion compensation from specially designed integrated-resonant unit elements, achromatic metalenses only work for one circular polarization can be designed [8,10]. Nevertheless, current achromatic metalenses almost are exclusively based on regular-shaped meta-atoms, such as cylinders or cuboids, depending on a few parameters to adjust the shapes [8,10,12,16]. The meta-atom designs are driven by human intuition and restricted to designers' experience. Metasurfaces with simple functions, like beam steering and chromatic focusing can be achieved through these regular-shaped meta-atoms, but it is getting more challenging for complex metasurface designs since it is difficult to find a set of satisfying meta-atoms from a small library. Although the meta-atoms library can be expanded by introducing more types of meta-atom shapes [11,17], like annular pillars and concentric pillars, it is still a very small portion of the entire design space. Furthermore, different types of shapes do not transit into each other smoothly, which causes discontinuities in the design space. What is worse, for the polarization-independent achromatic metalenses, the regular shapes suitable for the meta-atoms designs are quite limited due to the symmetry constrains. [11,17] In addition, those conventional design methods are usually time-consuming as it requires traversing all the parameters that control the meta-atom shape before final designs can be selected.

In order to open a larger design space and to design the meta-atoms without unintentional biases from prior human knowledge, here we developed an inverse design approach that creates random-shaped meta-atoms for broadband polarization-independent achromatic metalenses. Computer-aided inverse design methods have expedited the development of nanophotonic devices [18–26], such as particle accelerators [24], wavelength demultiplexers [20], and etc. Recently, optimizations and inverse design methods were implemented for designing high-performance metasurfaces [25,27–32]. Compared with the conventional intuition-based design methods, those new approaches improve considerably the device performance for multiple wavelengths responses [33], because they explore a much larger design space with free-form structures [21]. One popular inverse design method is gradient-descent-based optimization, such as topology optimization. It has been demonstrated for designing large-area metasurfaces [25,27,34,35], high-efficiency metasurfaces [25,36], elastic metasurfaces [37], and one-dimensional (1D) metalenses [38] However, it is a local optimizer that depends strongly on initial values, therefore multiple runs of optimization with different initial values are usually required in order to find a reasonable solution within a given design space [26]. In the past few years, data-driven optimization methods, such as deep neural network, have been applied for inverse design of nanophotonic structures including metasurfaces [19,22]. Although a trained neural network can rapidly design a device with the demanded properties, obtaining a large-size train set and training of the network are computationally expensive and time-consuming [19,22,23,26,28]. In addition, the neural network has to be re-trained if there are parameter changes that are not captured by the training set. Furthermore, non-uniqueness of the solution in inverse problems and the inaccuracies that may exist in the training set could lead to instable and nonconvergent training process[39].

Our new inverse design method solves those problems by combining a *fast random shape generator* and a *global optimization algorithm* (Fig. 1). We adopted the marching metaballs algorithm [40] for rapidly generating random shapes by morphing a two-dimensional (2D) binary round-cornered image continuously without abrupt changes (Fig. S1; see also Materials and Methods). This is been done by marching and merging a number of different-sized 'liquid' blobs (circles) in a designated framework. When two circles getting close to each other they will begin to merge like two liquid drops and similarly when two merged circles move further apart, they will become two separated circles. Compared with pixelized random shapes used in other inverse design methods, our algorithm provides vectorized shapes with much better resolution. It is nanofabrication-friendly as the produced shapes are inherently blobby with smooth boundaries. The limits in the nanofabrication can be automatically taken into account as the smallest feature sizes in the generated shapes are determined by the predefined smallest circles used in the algorithm. We coupled the shape generator with a surrogate-based optimizer [41,42] which searches for a global optimum using fewer cost function evaluations than most other global optimizers. [30]

Our inverse design method is inherently advantageous over conventional forward design methods and ideally can explore almost the entire design space for meta-atom designs. The method for the global optimum design does not require a good initial value or gradient information of the cost function. In addition, it minimizes the number of function calls to the computationally expensive full-wave evaluations.[30] We applied our inverse design method on unintuitively designing polarization-independent achromatic metalenses with random-shaped meta-atoms. The metalenses were numerically analyzed, nanofabricated, experimentally characterized, and they exhibit superior performance than that of the existing ones.

**Results**

**Inverse design of meta-atoms for achromatic metalenses**

Through an achromatic lens, light should focus on the same focal plane for different wavelengths within a range of interest. Therefore, the phase profile over the entire wavelength range should follow

$$\varphi(r,\omega) = -n\frac{\omega}{c} \cdot \sqrt{r^2 + F^2} + f(\omega) \tag{1}$$

where $n$ is the refractive index of the external environment where light is focused to, $\omega$ is the frequency of light, $r$ is the distance from the center of the metalens. $F$ is the focal length of the designed metalens. $f(\omega)$ is an arbitrary function that is only related to $\omega$. To satisfy this requirement, the phase responses of the meta-atoms need to form different phase profiles for different frequencies within the frequency range on an achromatic metalens.

In our design, we set the $f(\omega)$ in a general form $f(\omega) = \frac{\omega}{c} \cdot \sqrt{r_0^2 + F^2} + b_0$, and the required phase profile is

$$\varphi(r,\omega) = k_\varphi(r)(\omega - \omega_0) + \varphi_0(r) \tag{2}$$

where $k_\varphi(r) = -\frac{n}{c}\left(\sqrt{r^2 + F^2} - \sqrt{r_0^2 + F^2}\right)$, $\varphi_0(r) = -\frac{n\omega_0}{c}\left(\sqrt{r^2 + F^2} - \sqrt{r_0^2 + F^2}\right) + b_0$,

and $\omega_0$ is a reference frequency that can be chosen arbitrarily. $\varphi_0(r)$ is in fact the phase profile a chromatic lens designed to work at the reference frequency $\omega_0$. We can see that the required phase at any location $r$ is linearly dependent on the frequency $\omega$ since $k_\varphi(r) = \frac{d\varphi}{d\omega}$ and $\varphi_0(r) = \varphi(r, \omega_0)$ are both independent of $\omega$. One usual approach to get the needed meta-atoms is to obtain $k_\varphi(r)$ and $\varphi_0(r)$ from simulations for each of them and get the ones match the requirement. In this approach a pre-calculated meta-atom library is needed [11,12,17].

In our approach, it is not necessary to obtain a meta-atom library. We followed a multi-objective least squares formulation to define a cost function $C(r)$ shown below for our optimization

$$C(r) = w_\varphi \left\| \boldsymbol{\varphi}_{MA}(r) - k_\varphi(r)(\boldsymbol{\omega} - \omega_0) - \varphi_0(r) \right\|_2^2 + w_T \left\| 1 - \mathbf{T}_{MA}(r) \right\|_2^2 \quad (3)$$

where $w_\varphi$ and $w_T$ are predefined weighting parameters for adjusting the contributions from the phase error and the transmittance loss, respectively. For our design, we set $w_\varphi = w_T = 1$. $\|\cdot\|_2$ denotes $L_2$ norm. $\boldsymbol{\omega} = [\omega_1, \omega_2, \ldots, \omega_M]^T$ is a vector contains all the sampling frequencies, $\boldsymbol{\varphi}_{MA}(r)$ and $\mathbf{T}(r)$ are the vectors containing actual phase responses and transmittances, respectively, at the corresponding sampling frequencies from a meta-atom placed at position $r$. The cost function takes into account the distance of the phase from each sampling frequency to the required linear phase dispersion line as well as the reduction of transmittance from unity. The optimizer minimizes the cost function which ensures the resulting meta-atoms satisfy the phase and phase dispersion requirement over the entire range while having large transmittance.

The workflow of our inverse design algorithm is illustrated in Fig. 1. The meta-atoms in our design are invariant in the vertical direction which complies with our nanofabrication processes. We used the two-dimensional (2D) projection of the meta-atoms on the horizontal plane to indicate their shapes. In our algorithm, they are presented by the union of a set of circles with different center positions and radii [40] (Fig. S1; see also Materials and Methods). In order to have a set of reasonable initial meta-atoms for our optimizer, we first formed a chromatic metalens with cylindrical meta-atoms, which also work as a control set for comparison (Fig. S2; see also Materials and Methods). As the initial shape of the meta-atom is cylindrical, the initial circles are all positioned in the center of the unit-cell. The optimizer calculates the cost function value for the input shape with a full-wave-simulation-based evaluator, which extracts both the spectroscopic phase and transmittance information. If the cost function value is less than a predefined small number, the process stops, and the algorithm outputs the current best shape. Otherwise, the optimizer invokes the shape generation process, where the shape generator creates a set of new shapes by distorting the current shape through randomly marching the circles using the matching metaball algorithm [40]. We would like to emphasize that the surrogate algorithm we adapted greatly reduces the invocation times for the expensive (in time and computational resources) evaluator. Instead of directly calling the evaluator function to get full wave simulated spectrum information, a radial basis interpolation [43] is used to estimate the value of the cost function. Only the shape with lowest estimated cost function value will be evaluated by the evaluator. If the cost function value is reduced, the algorithm updates the record of the 'current best shape', otherwise, the record is kept the same as the previous one. The algorithm will iterate until the value of the cost function reduces to a predefined small value $\xi$. When the iteration shops, the algorithm outputs the 'current best shape' which has the minimized error in terms of phase, phase dispersion, and transmittance.

As a demonstration of our inverse design algorithm, we designed achromatic metalenses working in the near infrared whose working wavelengths range from 1.2 to 1.665 μm. It is worth noting that to have the resulting meta-atoms be insensitive to the input light polarization, we configured the shape generator to create shapes with four-fold reflectional symmetry by reflecting the resulting shape with respect to 0°, 45°, 90°, 135° four reflection axes (Fig. S1). Compared with metalens design based on geometrical phase, which only works for one circular-polarized light, our design is polarization-independent, making it more useful in practical applications. In a typical design process, as the number of iterations increases, we can see that the cost function value monotonically decreases (Fig. 2A), and the phase and phase dispersion of the meta-atom is getting closer to the required ones (Fig. 2 B and C). Simultaneously, its transmittance is getting closer to unity. The optimization stops when the cost function reaches a predefined small value, which in our case is 0.6. Finally, Two achromatic metalenses with different aperture size were created by placing the optimum meta-atoms in the corresponding positions, which provide required phase profiles for all the wavelengths within the designed range (Table S1).

**Simulation results**

We simulated our designed achromatic metalens consisting of random-shaped meta-atoms with the smaller aperture size (diameter 17.4 μm, focal length 15 μm, numerical aperture (NA) 0.5, shown in Fig. 3B) in a full-wave electromagnetic solver based on the finite different time domain (FDTD) method. As a comparison, we also simulated a metalens consisting of regular cylindrical meta-atoms with the same aperture size (Fig. 3A) without chromatic aberration correction.

From the realized phase profiles of the chromatic and achromatic metalenses at five evenly spaced frequency points across the designed frequency band (Fig. S3 B and D), we can see that the phase profile of the chromatic metalens matches the required one only at one frequency, $f = 180$ THz, while the phase profiles of the achromatic metalens show almost perfect match with the required ones at all the frequencies. This result is consistent with the result that the focal length of the chromatic metalens deviates from the designed value when the frequency moves away from 180 THz (Fig. 3C). In contrast, the focal length of the achromatic lens keeps unchanged, the same as designed, across the entire frequency range (Fig. 3D). This can be clearly seen from the intensity distributions for both chromatic metalens and achromatic metalens along the $z$-direction where $x = 0$ μm, $y = 0$ μm and intensity distribution along $x$-direction where $y = 0$ μm, $z = 15$ μm (Fig. S4). The simulations also show that our metalenses reveals consistently high focusing efficiency across the designed frequency range (Table S2).

**Characterization of the random shape achromatic metalens**

To further validate our achromatic metalens designs, we fabricated the larger-aperture metalens with focal length of 100 μm and NA of 0.2 (Fig. S5; see also the Device fabrication in Materials and Methods). The fabricated achromatic metalens (Fig. 4 B and C) was characterized by our home-built optical platform (Fig. 4A, also see the Measurement methods in Materials and Methods). We illuminate the metalens with a Gaussian beam of 1 mm waist, which was much larger than the metalens diameter to ensure uniform incident light. The light intensity profiles after the metalens were acquired by recording images at different distances from the metalens along the optical axis ($z$ axis).

The intensity profiles captured on two perpendicular planes ($yz$- and $zx$- planes) after the metalens clearly show tight focus spots at eight evenly spaced frequencies (from 188.75 THz to 250 THz) (Fig. 4 D and E). We note that our metalens can work at 180 THz, however that frequency is out of the range of our laser and hence it was not tested in our experiment. Quantitative focal length analysis shows that the lens has a focal length of about 100 μm consistently within the designed band (Fig. 5B). The intensity profiles on the focal plane reveal excellently diffraction-limited focal spots (Fig. 4F). Those results perfectly match our design parameters.

We also measured the focusing efficiency of our metalens. We followed the focusing efficiency defined in Ref. [5] which is the ratio of the measured optical power passing through the focal plane within a disk of three full width at half maximum (FWHM) dimeter to the measured input power. Our fabricated metalens exhibits the focusing efficiency of 41~50 % at all wavelengths (Fig. 5A), which is superior or comparable to other previously reported polarization-independent achromatic metalenses working in transmission mode (Table. S3 and Table S4). The standard deviation for our focusing efficiency across the whole working frequency band is only 3.28, which is much smaller than previously reported ones. In addition, we measured the focusing efficiency with different incidence polarizations by inserting an achromatic half-wave plate before the metalens. The measurement results show that the change of the focusing efficiency is negligible with different input polarizations, which validates the polarization insensitivity of our metalens. We define the longitudinal spot size $\delta F$ as the range around the focus where the normalized intensity is larger than 95% of the maximum value. Over the entire designed frequency range, the longitudinal spot size is small and covers the designed focal length.

To further analyze the performance of our metalens, we extracted the Strehl ratio from the measured intensity profiles. Strehl ratio is a parameter used to quantify lens aberrations, which is defined as the ratio between the peak of the point spread function (PSF) of a focusing system and that of an ideal aberrations-free lens's PSF. [10,11,25,44] The Strehl ratio of our achromatic metalens is consistently greater than 0.8 over the entire working frequency range (Fig. 5C), indicating the diffraction-limited performance of our metalens. Furthermore, the measured focus spot sizes at the focal plane are close to the theorical diffraction limited focus sizes (Fig. 5D).

## Discussion

In this work, we developed a new inverse design method that combines a fast random shape generation algorithm with a surrogate-based optimization, which can optimize the random-shaped meta-atoms to satisfy the complex phase and phase dispersion requirements.

We used it on designing random-shaped meta-atoms for metalenses with achromatic aberration compensation. Our metalenses show numerically and experimentally polarization-insensitive achromatic focusing working at a broad wavelength ranging from 1200 nm to 1665 nm. The measured focusing efficiency of our metalens is 41%~50% within the working band, which is superior or comparable to others. The standard deviation of the measured focusing efficiency is also much smaller than others, meaning the focusing efficiency variation is very small over the designed wavelength range. Strehl ratio larger than 0.8 over the entire working frequency range indicates the diffraction limited performance of our metalens. Furthermore, the measured focus spot sizes at the focal plane are close to the theorical diffraction limited focus sizes. All the longitudinal spot size within the designed frequency range covers the designed focal length, justifying the good agreement between the design and the realization.

In contrast to the conventional design method that rely on regular shapes and designers' intuition, our inverse design method reduced full-wave simulation times, since this method preclude the generation of the entire large meta-atom library. Using random shaped meta-atoms significantly expanded the design space. The random shapes generated from our marching metaball algorithm inherently have round edges and corners, which makes them nanofabrication-friendly and somewhat fabrication-error torrent. This design method also has the potential to design other metasurface-based devices, such as hyper-dispersion[45], large bandwidth beam steering metasurfaces, and polarization control metasurfaces. For different shape requirements based on need, such as rotational symmetry or mirror symmetry, our method can easily take into account the corresponding constraints on the generated random shapes.

In conclusion, our inverse-design method combining the random shape generation algorithm with global optimization solver is a powerful method for metasurface-based device design. We applied it on the design of achromatic metalenses with random shaped meta-atoms. Our simulation and experimental results show superior performance of our designed metalenses, working over a broad bandwidth of 465 nm in the near-infrared region. This novel method to correct the chromatic aberration can lead to great progress in compact optical systems, such as microscopes, digital cameras, and 3D displays. We envision that this method can relieve people from tedious processes of trial and be a possible solution for high-dimensional problems.

**Materials and methods**

**Random shape generation.** For the random shape generation, we used the technique for rendering metaballs, which was invented by Jim Blinn in the early 1980s.[40] First, a build function was set up based on a total number of $N$ circle's radii and positions (Fig. S1A).

$$f(x, y) = \sum_{i=0}^{N} \frac{r_i^2}{(x-x_i)^2 + (y-y_i)^2} \tag{4}$$

Then we can get contour lines at different heights. Finally, selecting a certain contour, we can get an arbitrary blobby shape as shown in Fig. S1B. To make this random shape has four-fold reflectional symmetry, we can reflect the shape four times with respect to 0°, 45°, 90°, 135° reflection axes. The build function therefore becomes

$$f_{\text{four-fold}}(x, y) = f(x, y) + f(-x, y) + f(x, -y) + f(-x, -y)$$
$$+ f(y, x) + f(-y, x) + f(y, -x) + f(-y, -x) \tag{5}$$

In Fig. S1C, the yellow shape is from the original build function, and the green one is from the altered build function with four-fold reflectional symmetry. The shape has smooth boundaries, which is friendly for fabrication. Other than that, we also restrict the shape to have a minimum wall thickness and a minimum gap size according to the nanofabrication limits. Vectorized images (polygons in our case) were used for storing the generated random shapes to achieve better image resolution and scalability than bitmap images.

**Initial chromatic metalens design**. The initial design is a chromatic metalens consists of cylinder meta-atoms (Fig. S2A). To design a metalens, the first step is to find a reasonable configuration of a unit-cell, such as the period and the height of the antenna material. Generally, the larger the period and/or height is, the larger the phase range the building block can cover. However, to suppress the high-order diffraction, the period cannot be very large and was chosen to be 600 nm in our designs. Considering the nanofabrication constraints, the minimum radius we set was 80 nm and the maximum aspect ratio (width : height) of Si antennas was chosen to be 1:10. As shown in Fig. S2B, by varying the duty cycle, which defined as the cylinder diameter over the unit-cell period, the phase shift for light transmitted through the meta-atom can cover $2\pi$. Based on equation (1), we designed a chromatic metalens working at 180 THz frequency with a focal length of 15 µm (Fig. S2D). This chromatic metalens design was used as the initial values for our optimizer and also as a control sample for comparison.

**Device fabrication.** The metalens samples were fabricated based on top-down approach for in-plane nanostructures. First, 800 nm thick amorphous silicon was deposited on a diced fused silica substrate with plasma-enhanced chemical vapor deposition (PECVD). Second, sample was processed by oxygen plasma and Surpass 3000 to promote adhesion between silicon and the electron-beam (e-beam) resist film. 200-nm-thick ZEP-520A e-beam resist was spin-coated followed by thermally evaporation of a 20-nm-think Au film as a charge conducting layer before e-beam exposure. The pattern was exposed by an e-beam writer (Raith EBPG5200) and immersed in N-ayml-acetate for 3 minutes to develop. Then a 90-nm-thick $Al_2O_3$ film was deposited by e-beam evaporation as a hard mask for etching. The patterns were transferred to $Al_2O_3$ by a standard lift-off process in heated N-methylpyrrolidone-based solution (Microdeposit remover 1165). Next, the sample was dry-etched in an inductively-coupled-plasma reactive ion etching (ICP-RIE) system (Alcatel Speeder 100) with $C_4F_8$ / $SF_6$ to define the vertical profile of the nanoantennas. Finally, the $Al_2O_3$ hard mask was removed in $NH_4OH:H_2O_2:H_2O$ (1:1:1) solution. (Fig. S5)

**Measurement methods.** A tunable Ti:Sapphire laser with optical parametric oscillator (OPO) was used as the input light source. The output of the laser was linear polarized. A half-wave plate was placed before the metalens to control the input polarization. Light passing through the metalens was captured by an objective lens. A tube lens after the objective lens was used to form a image on a near-infrared camera. The intensity distribution on the xy-plane can be captured directly by the camera (assuming light was propagating along the z-axis). In order to visualize the intensity distribution on the xz- and yz- planes, we recorded a series of intensity distributions in the xy-plane at different z positions. Regarding the efficiency measurement, the input power was measured through a precision tunable iris whose aperture was tuned to be the same size as the metalens. The transmitted power was measured at the plane immediately after the metalens. The focused power was measured through an iris of which the diameter equaled to three times of FWHM of focus spot placed at focal plane of the metalens.

## Acknowledgements

**Funding:** The work was partially supported by the Moore Inventor Fellow award from the Gordon and Betty Moore Foundation, the National Aeronautics and Space Administration Early Career Faculty Award (NASA ECF) under grant no. 80NSSC17K0528, the Office of Naval Research (ONR) Basic Research Challenge (BRC) under grant no. N00014-18-1-2371, the National Institute on Aging of the National Institutes of Health under award no. 1R56AG062208-01, and the Penn State MRSEC, the Center for Nanoscale Science, under grant no. NSF DMR-1420620.

## Contributions

X.N. conceived and supervised the project. X.J.Z., H.Y.H. and S.Y.C developed random shape generator. X.J.Z., X.W.Z. and X.C. performed simulations. X.G. and Y.D. fabricated the metalenses. X.J.Z., X.X.G., X.W.Z. and Y.M.D characterized the metalenses. X.J.Z., H.Y.H., X.G., X.W.Z. and X.N. wrote the manuscript. All authors discussed the results and commented on the manuscript.

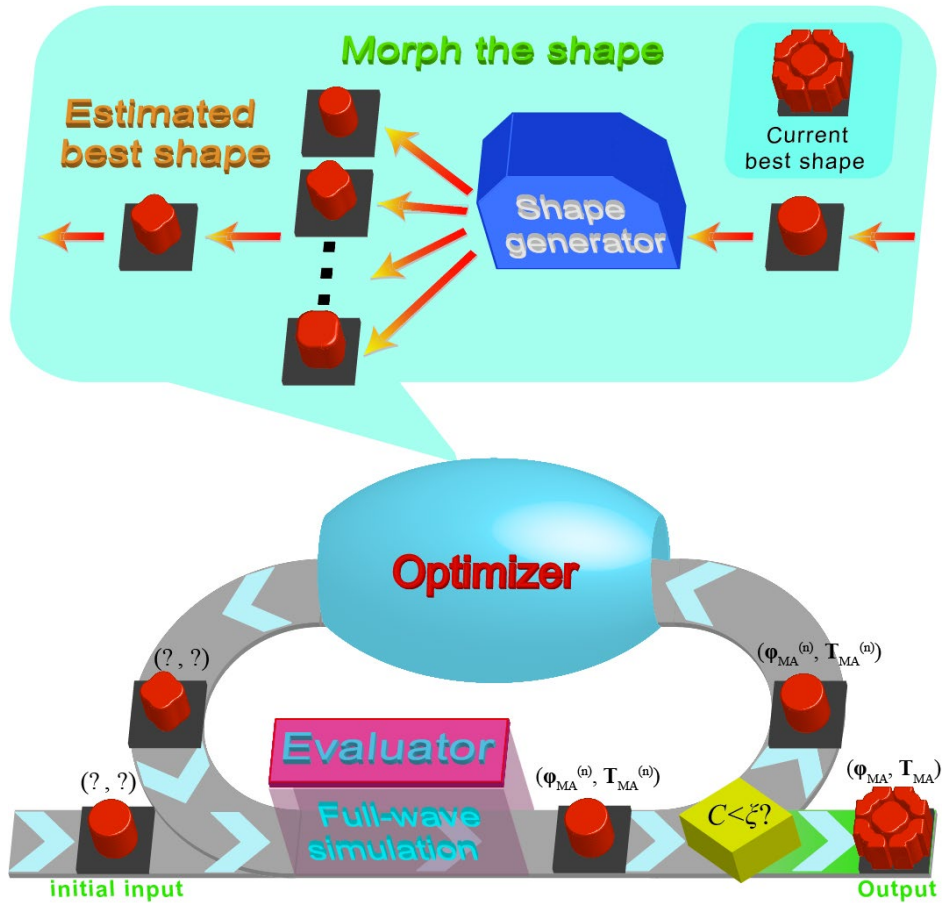

**Fig. 1. The inverse design workflow for designing random-shaped meta-atoms**. After taking in an initial meta-atom shape, the evaluator calculates its phases and transmittances, ($\varphi$, **T**), at all frequency points within the demanded frequency regime through full-wave simulation. Then the shape will be sent to the optimizer where the embedded shape generator creates a set of new shapes by morphing the current shape through marching the circles using the matching metaball algorithm. A surrogate model based on radial basis interpolation is established and used to estimate the cost function value $C$ for the morphed shapes. The optimizer then outputs the estimated best shape among the morphed ones and sends it to the evaluator to calculate its accurate phases and transmittances. After that, the cost function value $C$ is evaluated and compared with a predefined

threshold $\xi$. If the condition $C < \xi$ is fulfilled, the algorithm outputs the best shape. Otherwise, algorithm begins the next iteration, and the shape replaces the 'current best shape' record if its cost function value is lower than the previous one.

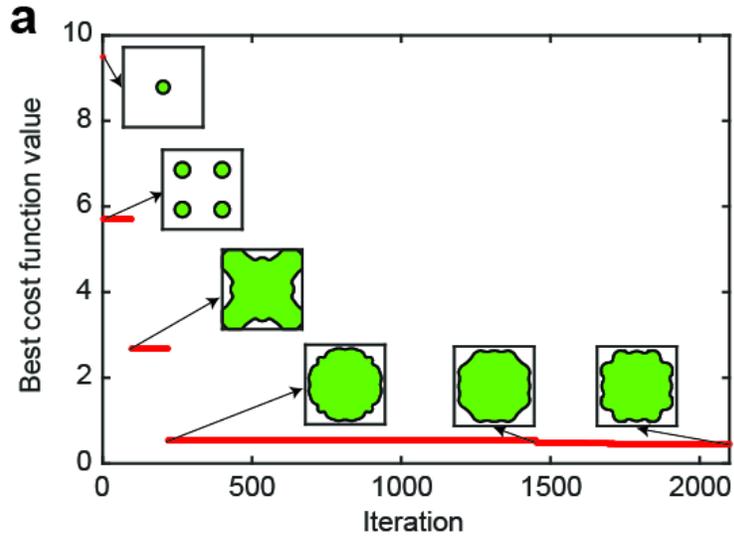

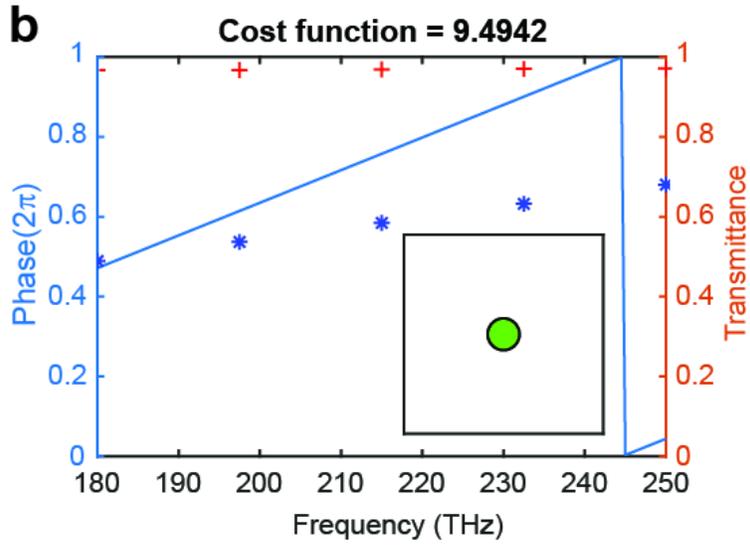

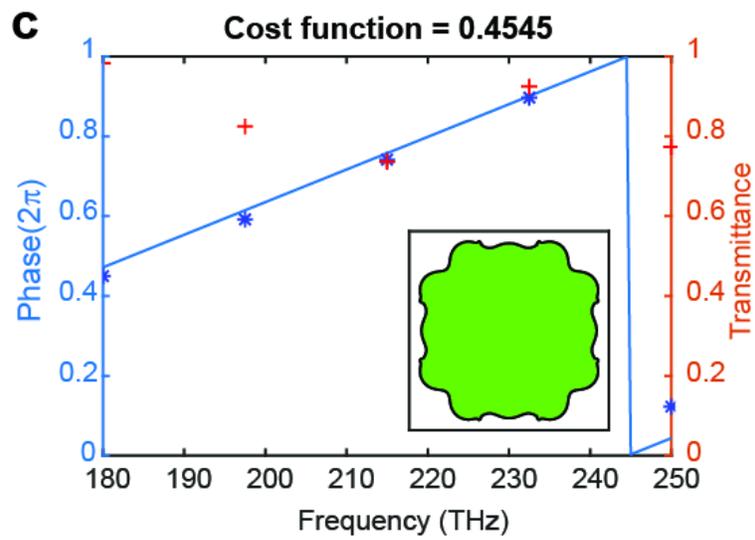

**Fig. 2. A typical optimization process for a random-shaped meta-atom.** (**a**) The recorded cost function values of the 'current best shape' at each iteration. The insets show the corresponding 'current best shapes.' (**b**) and (**c**) The full-wave simulated phase and transmittance for (b) the initial meta-atom shape (shown in the insets) and (c) the optimized meta-atom shape. The shapes are shown in the insets. Their corresponding cost function values are 9.4942 and 0.4545, respectively. The blue solid line is the required phase for an achromatic metalens design at all frequencies. The blue asterisks and red pluses are the phase and transmittance provided by the meta-atom, respectively. We can clearly see that the phase of the meta-atom is close to the required phase at all frequencies while maintaining relatively high transmittance after optimization.

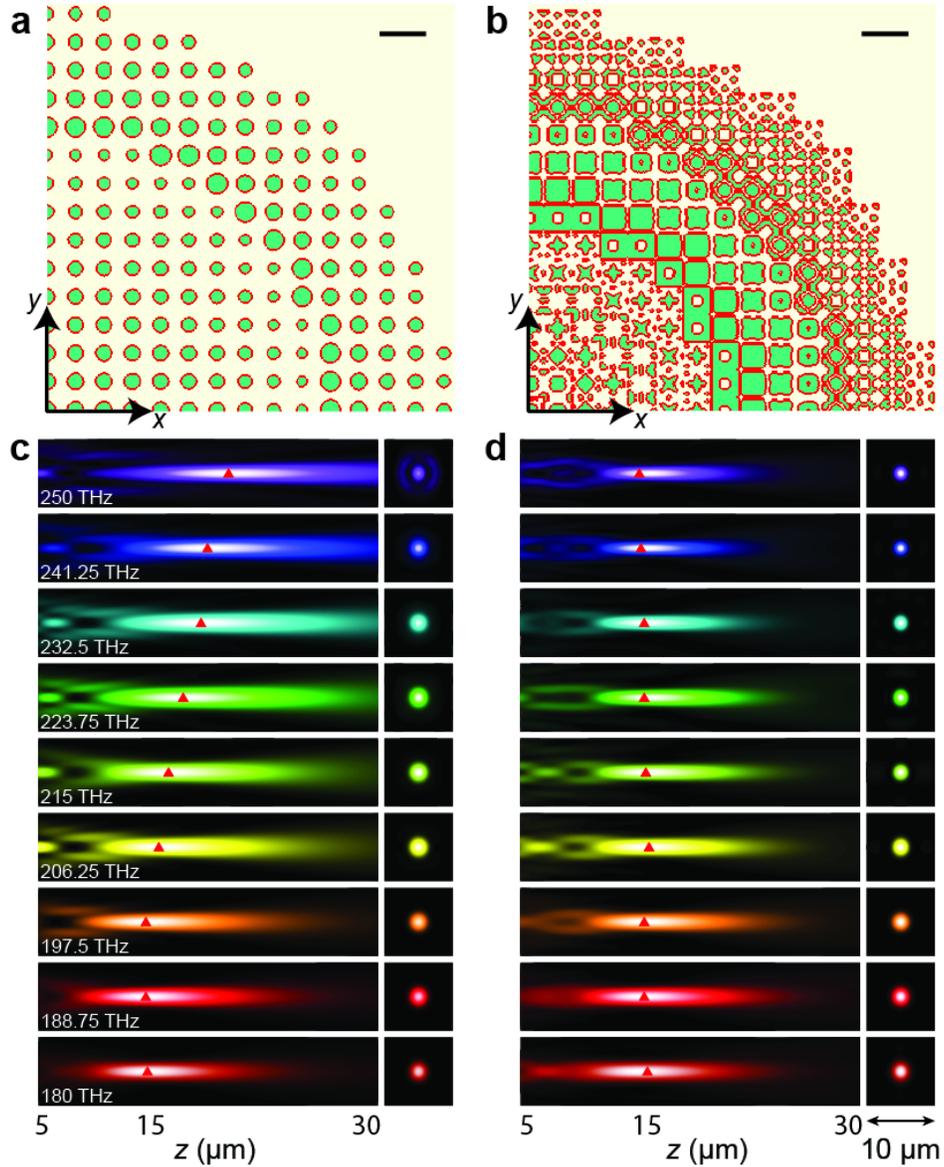

**Fig. 3. Simulated intensity profiles for the chromatic and achromatic metalenses.** (**a**) and (**b**) A quarter of (a) the chromatic metalens design and (b) the achromatic metalens design. The scale bar is 1 μm. (**c**) and (**d**) The intensity profiles in a longitudinal cross section (*yz*-plane) at $x = 0$ μm (left-side panels) and the focal planes (*xy*-planes going through the focal points) (right-side panels) obtained from full-wave simulations of (c) the designed chromatic metalenses and (d) the designed achromatic metalens at nine evenly spaced frequencies. The red triangles indicate the focal points.

The focal lengths of the chromatic metalens gets larger with the higher frequencies. In contrast, those of the achromatic metalens are approximately the same for all frequencies. The tested frequencies are evenly spaced from 180 THz to 250 THz (i.e. 180 THz, 188.75 THz, 197.5 THz, 206.25 THz, 215 THz, 223.75 THz, 232.5 THz, 241.25 THz, and 250 THz) from bottom to top panels.

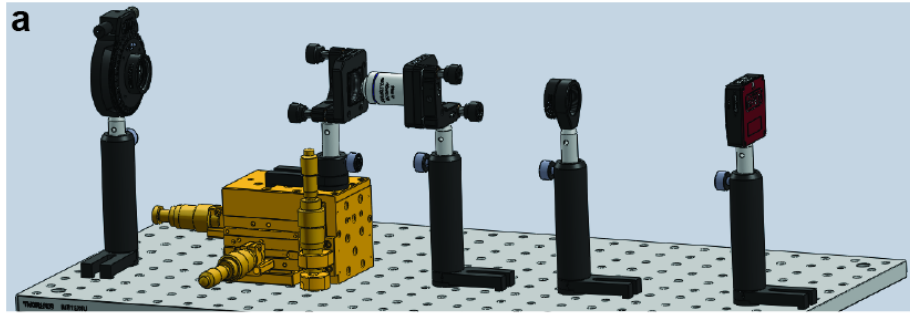
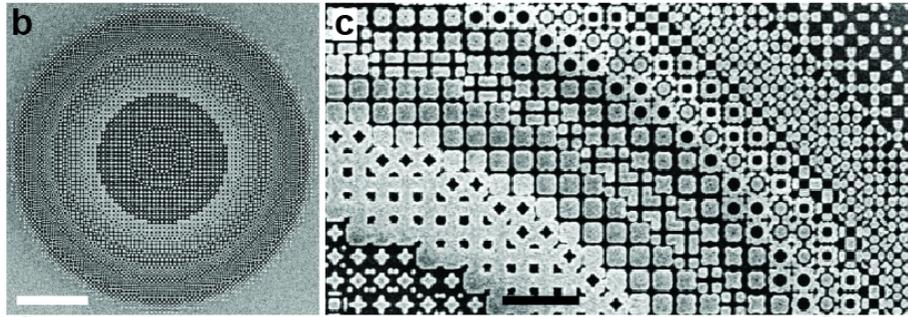
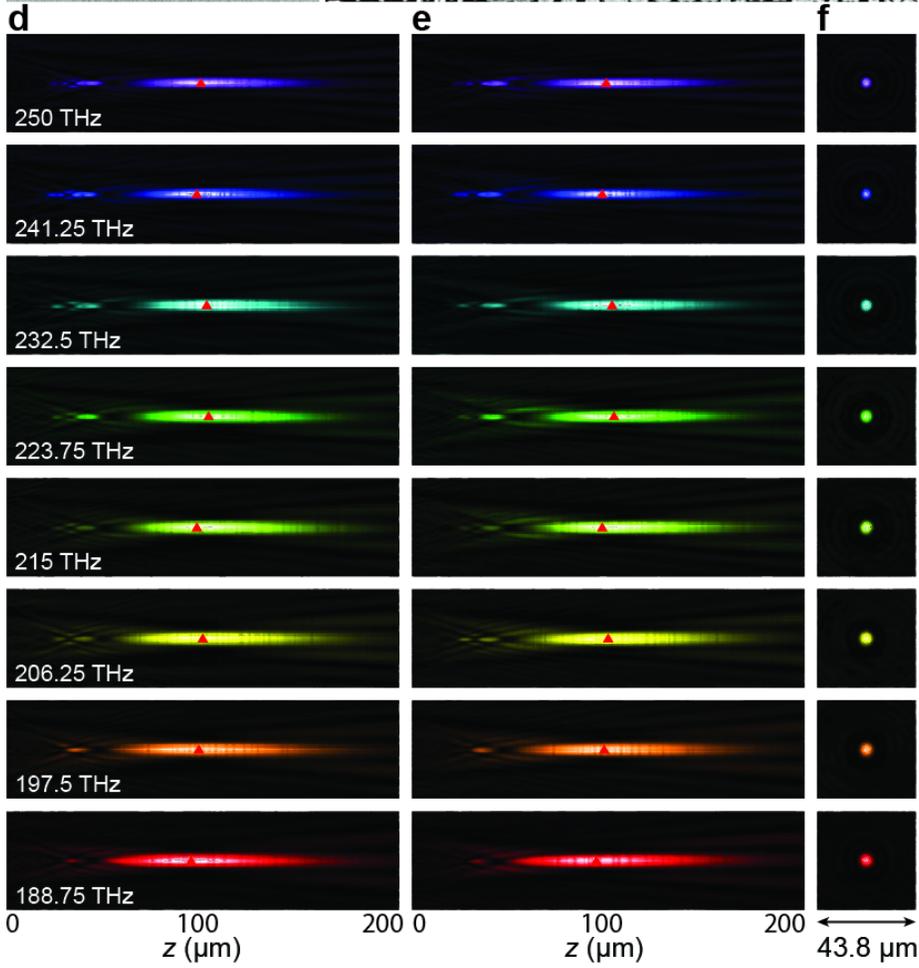

**Fig. 4: Experimental results of a fabricated achromatic metalens with random-shaped meta-atoms.** (**a**) The homebuilt optical setup used for metalens characterization (only showing the key components). Collimated incident light from a tunable laser passes through a half-wave plate for controlling the input polarization before impinging onto the metalens. Light passing through the metalens was captured by an objective. The objective has larger NA than the metalens to ensure all the transmitted light being collected. The objective together with a tube lens after it form a 4-F system and project the image on a near-infrared camera. The intensity profiles after the metalens at different distances ($z$ positions) were acquired by moving the metalens along the light propagation direction ($z$ direction). (**b**) and (**c**) Scanning electron microscopy images of a fabricated achromatic metalens with focal length $f = 100$ μm. The scale bars are 10 μm in (b) and 2 μm in (c). (**d**) and (**e**) The measured intensity profiles in two longitudinal cross sections, (d) the $xz$-plane ($y = 0$) and (e) the $yz$-plane ($x = 0$), of the fabricated achromatic metalens at eight evenly spaced frequencies. The red triangles reveal the largest intensity points in the intensity profiles, indicating the focal points. The measured focal lengths of the achromatic metalens are approximately the same for all frequencies, validating its achromatic performance. (**f**) The measured intensity profiles at the focal planes ($xy$-planes going through the focal points) of the achromatic metalens at eight evenly spaced frequencies. The measured frequencies are evenly spaced from 188.75 THz to 250 THz (i.e. 188.75 THz, 197.5 THz, 206.25 THz, 215 THz, 223.75 THz, 232.5 THz, 241.25 THz, and 250 THz) from bottom to top panels. Note that 180 THz out of the output range of our laser and thus not tested.

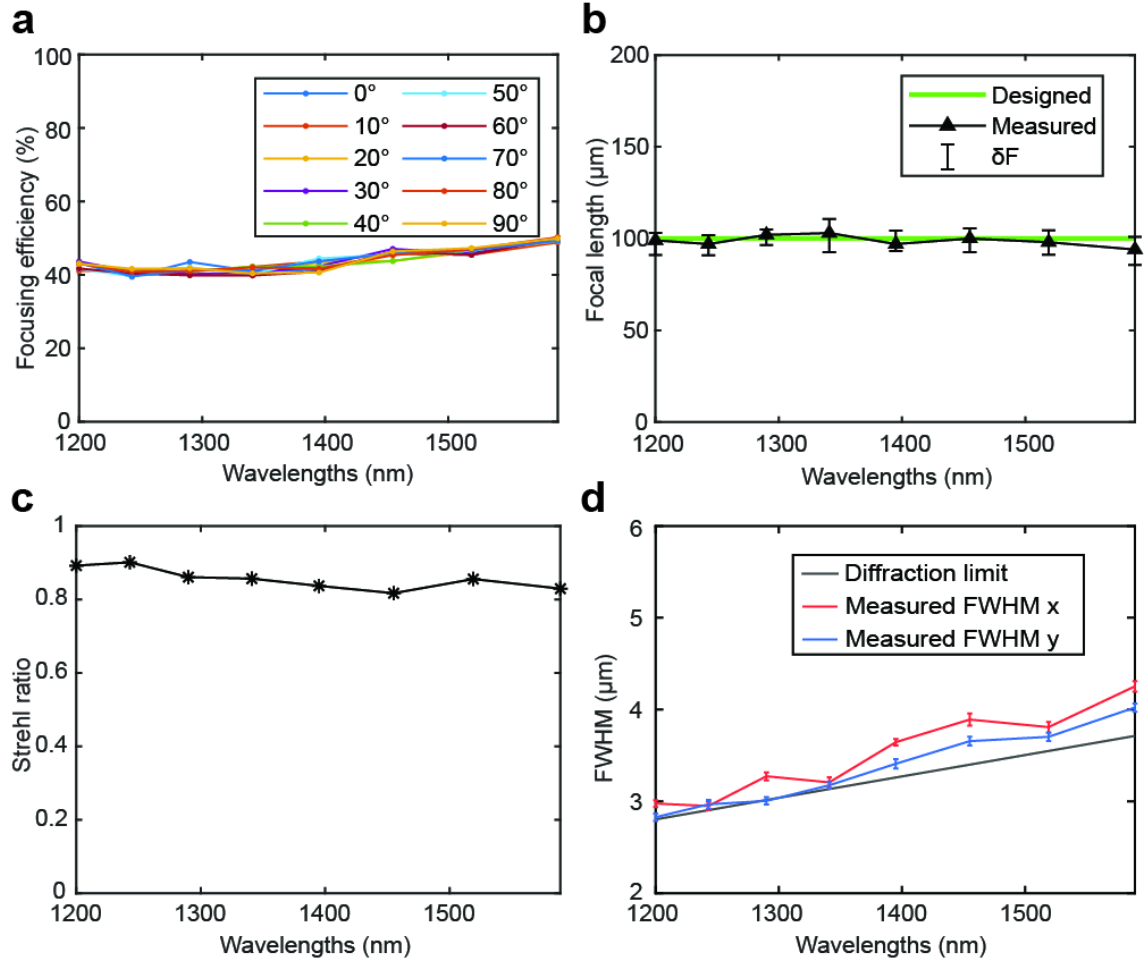

**Fig 5: Measured properties of the achromatic metalens for quantitative performance analysis.** (**a**) Focusing efficiency of the metalens with different input linear polarization. (**b**) The focal lengths at different frequencies extracted from the intensity profiles. The green line indicates the designed focal length, 100 μm. The black triangles indicate the measured ones. The error bars show the *δF*, which is defined as the range where the intensity is larger than 95% of the maximum. (**c**) Strehl ratio of the achromatic metalens extracted from the measured intensity profiles. It is defined as the ratio between the peak of the point spread function (PSF) of our metalens and that of an ideal aberrations-free lens's PSF. The Strehl ratio of our achromatic metalens is consistently greater than 0.8 over the entire working frequency range, indicating the diffraction-limited

performance. (**d**) The measured FWHM along the *x* direction (red solid line) and the *y* direction (blue solid line) of intensity profiles at the focal plane of the achromatic metalens. The black solid line indicates that from an ideal diffraction-limited lens.